\documentclass{article}
\usepackage{spconf,amsmath,epsfig}

\title{Data Driven Discovery in Astrophysics}
\name{ Giuseppe Longo$^1$, Massimo Brescia$^2$, S.G. Djorgovski$^3$, Stefano Cavuoti$^2$, Ciro Donalek$^3$\thanks{longo@na.infn.it}}
\address{1-Department of Physics, University Federico II, via Cintia 6, Napoli, Italy\\
2-INAF-Astronomical Observatory of Capodimonte, via Moiariello 16, Napoli, Italy\\
3-Center for Data Driven Discovery, California Institute of Technology, Pasadena, 90125- CA, USA}

\begin{document}
%
\maketitle
\begin{abstract}
We review some aspects of the current state of data-intensive astronomy, its methods, and some outstanding data analysis challenges.
Astronomy is at the forefront of “big data” science, with exponentially growing data volumes and data rates, and an ever-increasing complexity, 
now entering the Petascale regime. 
Telescopes and observatories from both ground and space, covering a full range of wavelengths,  feed the data via processing pipelines into 
dedicated archives, where they can be accessed for scientific analysis. 
Most of the large archives are connected through the Virtual Observatory framework, that provides interoperability standards and services, and 
effectively constitutes a global data grid of astronomy. 
Making discoveries in this overabundance of data requires applications of novel, machine learning tools. 
We describe some of the recent examples of such applications.
\end{abstract}
\section{Introduction}\label{sec:intro}
Like most other sciences, astronomy is being fundamentally transformed by the Information and Computation Technology (ICT) revolution. 
Telescopes both on the ground and in space generate streams of data, spanning all wavelengths, from radio to gamma-rays, and non-electromagnetic 
windows on the universe are opening up: cosmic rays, neutrinos, and gravitational waves. 
The data volumes and data rates are growing exponentially, reflecting the growth of the technology that produces the data. 
At the same time, we see also significant increases in data complexity and data quality as well. This wealth of data is greatly accelerating our
understanding of the physical universe.

It is not just the data abundance that is fueling this ongoing revolution, but also Internet-enabled data access, and data re-use. 
The informational content of the modern data sets is so high as to make archival research and data mining not merely profitable, but practically 
obligatory: in most cases, researchers who obtain the data can only extract a small fraction of the science that is enabled by it. 
Furthermore, numerical simulations are no longer just a crutch of an analytical theory, but are increasingly becoming the dominant or even the 
only way in which various complex phenomena (e.g., star formation or galaxy formation) can be modeled and understood. 
These numerical simulations produce copious amounts of data as their output; in other words, theoretical statements are expressed not as formulae, 
but as data sets. 
Since physical understanding comes from the confrontation of experiment and theory, and both are now expressed as ever larger and more complex data 
sets, science is truly becoming data-driven in the ways that are both quantitatively and qualitatively different from the past.
The situation is encapsulated well in the concept of the “fourth paradigm” \cite{Hey2009}, adding to experiment, analytical theory, and numerical 
simulations as the four pillars of modern science. 
This profound, universal change in the ways we do science has been recognized for over a decade now, sometimes described as e-Science, cyber-science, 
or cyber-infrastructure.

\section{Data Overabundance, Virtual Observatory and Astroinformatics}\label{sec:overabundance}
A confluence of several factors pushed astronomy to the forefront of data-intensive science. 
The first one was that astronomy as a field readily embraced, and in some cases developed, modern digital detectors, such as the CCDs or digital 
correlators, and scientific computing as a means of dealing with the data, and as a tool for numerical simulations.
The culture of e-Science was thus established early (circa 1980s), paving the way for the bigger things to come. 
The size of data sets grew from Kilobytes to Megabytes, reaching Gigabytes by the late 1980s, Terabytes by the mid-1990s, and currently 
Petabytes (see Fig. 1). 
Astronomers adopted early universal standards for data exchange, such as the Flexible Image Transport System (FITS; \cite{wells1981}).

The second factor, around the same time, was the establishment of space missions archives, mandated by NASA and other space agencies, 
with public access to the data after a reasonable proprietary period (typically 12 to 18 months). 
This had a dual benefit of introducing the astronomical community both to databases and other data management tools, and to the culture of 
data sharing and reuse. 
These data centers formed a foundation for the subsequent concept of a Virtual Observatory \cite{hanish2001}.
The last element was the advent of large digital sky surveys as the major data sources in astronomy. 
Traditional sky surveys were done photographically, ending in 1990s; those were digitized using plate-scanner machines in the 1990s, thus 
producing the first Terabyte-scale astronomical data sets, e.g., the Digital Palomar Observatory Sky Survey (DPOSS; \cite{GSD1999}). 
They were quickly superseded by the fully digital surveys, such as the Sloan Digital Sky Survey (SDSS; \cite{york2000}), 
and many others (see, e.g. \cite{GSD2012c} for a comprehensive review and references).
Aside from enabling a new science, these modern sky surveys changed the social psychology of astronomy: traditionally, 
observations were obtained (and still are) in a targeted mode, covering a modest set of objects, e.g., stars, galaxies, etc. 
With modern sky surveys, one can do first-rate observational astronomy without ever going to a telescope. 
An even more powerful approach uses data mining to select interesting targets from a sky survey, and pointed observations 
to study them in more detail.

This new wealth of data generates many scientific opportunities, but poses many challenges as well: how to best store, 
access, and analyze these data sets, that are several orders of magnitude larger than what astronomers are used to do on their desktops? 
A typical sky survey may detect $\sim 10^8 - 10^9$ sources (stars, galaxies, etc.), with $\sim 10^2 - 10^3$ 
attributes measured for each one. Both the scientific opportunities and the technological challenges are then amplified by data fusion, across
different wavelengths, temporal, or spatial scales.

\subsection{Virtual Observatory}\label{virtualobs}
The Virtual Observatory (VO, \cite{brunner2001a,GSD2002a,hanish2001}) was envisioned as a complete, distributed (Web-based) research environment for 
astronomy with large and complex data sets, by federating geographically distributed data and computing assets, and the necessary 
tools and expertise for their use. 
VO was also supposed to facilitate the transition from the old data poverty regime, to the regime of overwhelming data abundance, 
and to be a mechanism by which the progress in ICT can be used to solve the challenges of the new, data-rich astronomy. 
The concept spread world-wide, with a number of national and international VO organizations, now federated through the International 
Virtual Observatory Alliance (IVOA; http://ivoa.net).
One can regard the VO as an integrator of heterogeneous data streams from a global network of telescopes and space missions, enabling data
access and federation, and making such value-added data sets available for a further analysis. 
The implementation of the VO framework over the past decade was focused on the production of the necessary data infrastructure, 
interoperability, standards, protocols, middleware, data discovery services, and a few very useful data federation and analysis services 
(see \cite{hanish2007, graham2007}, for quick summaries and examples of practical tools and services implemented
under the VO umbrella).

Most astronomical data originate from sensors and telescopes operating in some wavelength regime, in one or more of the following forms:
images, spectra, time series, or data cubes. 
A review of the subject in this context was given in \cite{brunner2001b}. 
Once the instrumental signatures are removed, the data typically represent signal intensity as a function of the position on the sky, 
wavelength or energy, and time. 
The bulk of the data are obtained in the form of images (in radio astronomy, as interferometer fringes, but those are also converted 
into images). 

\begin{figure}[htb]
\begin{minipage}[b]{1.0\linewidth}
\centerline{\epsfig{figure=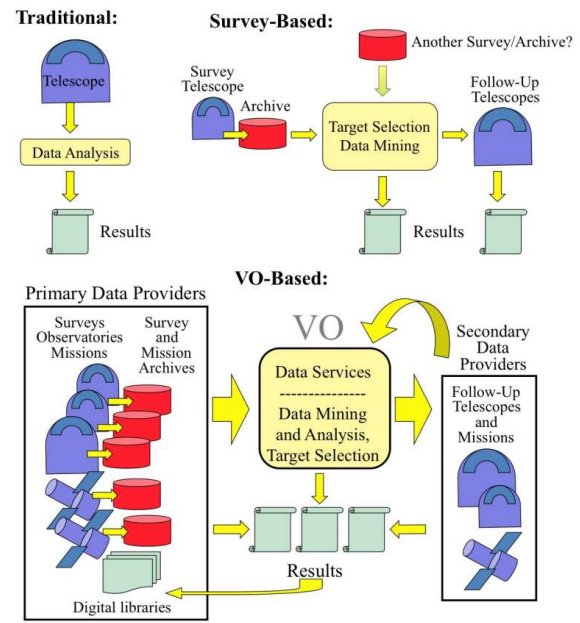,width=8.0cm}}
\caption{The evolving modes of observational astronomy. Top left: 
In the traditional approach, targeted observations from a single telescope (sensor), sometimes
combined with other data, are used to derive science. This mode is typical of Megabyte
to Gigabyte-scale data sets. Top right: In the survey mode, data from a given survey
are stored in an archive, and may be used to produce science on its own. Sometimes,
they may be matched to another survey. Selection of interesting targets using data
mining can then lead to new targeted observations, and new results. This mode is
characterized by Terabyte scale data sets. Bottom: In the VO mode, a large variety of
surveys, space missions, and ground-based observatory archives are federated in the
VO framework. Data fusion can lead to new science, or can be used to select targets
for follow-up observations, that themselves contribute to the evolving data grid. This
mode is characteristic of Terabyte to Petabyte-scale data sets. A new generation of
synoptic sky surveys imposes a requirement that the data-to-research cycle happens
in a real time. In practice, all three modes continue to coexist.}
\end{minipage}
\end{figure}

The sensor output is then processed by the appropriate custom pipelines, that remove instrumental signatures and perform calibrations.
In most cases, the initial data processing and analysis segments the images into catalogs of detected discrete sources (e.g., stars, galaxies,
etc.), and their measurable attributes, such as their position on the sky, flux intensities in different apertures, morphological descriptors of the
light distribution, ratios of fluxes at different wavelengths (colors), and so on. 
Scientific analysis then proceeds from such first-order data products.
In the case of massive data sets such as sky surveys, raw and processed sensor data, and the initial derived data products such as source catalogs
with their measured attributes are provided through a dedicated archive, and accessible online.

The Virtual Observatory (VO) framework aims to facilitate seamless access to distributed heterogeneous data sets, for example, combining
observations of the same objects from different wavelength regimes to understand their spectral energy distributions or interesting correlations among their properties. 
The International Virtual Observatory Alliance (IVOA) is charged with specifying the standards and protocols
that are required to achieve this. 
A common set of data access protocols ensures that the same interface is employed across all data archives, no
matter where they are located, to perform the same type of data query.
Although common data formats may be employed in transferring data, individual data providers usually represent and 
store their data and metadata in their own way. 
Common data models define the shared elements across data and metadata collections and provide a framework for describing relationships between them so
that different representations can interoperate in a transparent manner.
Most of the data access protocols have an associated data model, e.g., the Spectral data model defines a generalized model for spectrophotometric 
sequences and provides a basis for a set of specific case models, such as Spectrum, SED and TimeSeries. 
There are also more general data models for spatial and temporal metadata, physical units, observations and their provenance, and characterizing 
how a data set occupies multidimensional physical space.
When individual measurements of arbitrarily named quantities are reported, either as a group of parameters or in a table, their broader
context within a standard data model can be established through the IVOA Utypes mechanism. 
Namespaces allow quantities/concepts defined in one data model to be reused in another one.
Data models can only go so far in tackling the heterogeneity of data sources since they provide a way to identify and refer to common elements but
not to describe how these are defined or related to each other. 
Concept schemes, from controlled vocabularies to thesauri to ontologies, specify in increasing levels of detail the domain knowledge that is 
ultimately behind the data models. 
It then becomes possible, for example, to automatically construct a set of extragalactic sources with consistent distances, even if each initially 
has it specified in a different way; the Tully-Fisher relationship can be used with those with HI line widths whereas surface brightness and velocity 
dispersion can be used for elliptical galaxies.
Finally, the IVOA provides a Registry tool where descriptions of available data archives and services can be found, e.g., catalogs of white dwarfs or 
photometric redshift services.

\subsection{Beyond VO: Astroinformatics}\label{astroinfo}

While much still remains to be done, data discovery and access in astronomy have never been easier, and the established structure can at least 
in principle expand and scale up to the next generation of sky surveys, space missions, etc.
What is still lacking is a powerful arsenal of widely available, scalable tools needed to extract knowledge from these remarkable data sets.
The key to further progress in this area is the availability of data exploration and analysis tools that can operate on the Terascale data
sets and beyond. Progress in this arena is being made mainly by individual research groups in universities, or associated with particular 
observatories and surveys. 

Thus we now have an emerging field of Astroinformatics, a bridge field between astronomy on one side, and ICT and applied CS on the other 
(see, e.g., \cite{borne2009}). The idea behind Astroinformatics is to provide an informal, open environment for the exchange of ideas, software, 
etc., and to act as a “connecting tissue” between the researchers working in this general arena.
The motivation is to engage a broader community of researchers, both as contributors and as consumers of the new methodology for data-intensive
astronomy, thus building on the data-grid foundations established by the VO framework. 

A good introduction to Astroinformatics are the talks and discussions at the series of the international Astroinformatics conferences, starting
with http://astroinformatics2010.org.

\section{Data Mining and Knowledge Discovery}\label{sec:datamining}
Data – no matter how great – are just incidental to the real task of
scientists, knowledge discovery. Traditional methods of data analysis
typically do not scale to the data sets in the Terascale regime, and/or
with a high dimensionality. 
Thus, adoption of modern data mining (DM) and Knowledge Discovery in Databases (KDD) techniques becomes a 
necessity. Large data volumes tend to preclude direct human examination
of all data, and thus an automatization of these processes is needed, requiring use of Machine Learning (ML) techniques. 
Astronomical applications of ML are still relatively recent and restricted to a handful of problems. This is surprising, given the data
richness and a variety of possible applications in the data-driven astronomy. 
Sociological challenges aside, there are
some technical ones that need to be addressed.
First, a large family of ML methods (the so called supervised ones) requires the availability of relatively large and 
well characterized knowledge bases (KB), e.g., reliable (“ground truth”) training data sets of examples from which the 
ML methods can learn the underlying patterns and trends. 
Such KBs are relatively rare and are available only for a few specific problems.
Second, most ML algorithms used so far by the astronomers cannot deal well with missing data (i.e., no measurement was obtained for a
given attribute) or with upper limits (a measurement was obtained, but
there is no detection at some level of significance). 
While in many other fields (e.g., market analysis and many bioinformatics applications) this
is only a minor problem since the data are often redundant and/or can
be cleaned of all records having incomplete or missing information, in
astronomy this is usually not so, and all data records, including those
with an incomplete information, are potentially scientifically interesting
and cannot be ignored.

Finally, scalability of algorithms can be an issue. Most existing ML
methods scale badly with both increasing number of records and/or of
dimensionality (i.e., input variables or features): the very richness of
our data sets makes them difficult to analyze. 
This can be circumvented by extracting subsets of data, performing the training 
and validation of the methods on these manageable data subsets, and then extrapolating
the results to the whole data set. This approach obviously does not use
the full informational content of the data sets, and may introduce biases
which are often difficult to control. 
Typically, a lengthy fine tuning procedure is needed for such sub-sampling experiments, which may require
tens or sometimes hundreds of experiments to be performed in order to
identify the optimal DM method for the problem in hand, or, a given
method, the optimal architecture or combination of parameters.

Examples of uses of modern ML tools for analysis of massive astronomical data sets include:
automated classification of sources detected in sky surveys as stars (i.e., unresolved) vs. galaxies (resolved morphology) using Artificial Neural Nets (ANN) or Decision Trees (DT)
(e.g., \cite{weir1995, odewan2004, donalek2006}).
Brescia et al. \cite{brescia2012} have recently used a ML method for a different
type of resolved/unresolved objects separation, namely the identification
of globular clusters in external galaxies.
Another set of ML applications is in classification or selection of objects of a given type in some parameter space, e.g., colors. 
This is particularly well suited for the identification of quasars and other active
galactic nuclei, which are morphologically indistinguishable from normal stars, but represent vastly different physical phenomena (\cite{dabrusco2009,dabrusco2012,richards2009}).
Yet another application is estimates of photometric redshifts, that are derived from colors 
rather than from spectroscopy (\cite{tagliaferri2002, firth, hildebrandt2010, cavuoti2012}). Laurino et al. \cite{laurino2011} implemented a hybrid procedure based
on a combination of unsupervised clustering and several independent
classifiers that has improved the accuracy, for both normal galaxies and
quasars.

The rapidly developing field of time-domain astronomy poses some new challenges.  A new generation of synoptic sky surveys produces data streams that correspond to the traditional, one-pass sky surveys many times repeatedly \cite{GSD2012c,GSD2012a,GSD2013a,GSD2012d,GSD2011e}.  
In addition to the dramatic increase of data rates and the resulting data volumes, and all of the challenges already posed by the single-pass sky surveys, there is a need to identify, characterize, classify, and prioritize for the follow-up observations any transient events or highly variable sources that are found in the survey data streams.  
Since many such events are relatively short in duration, this analysis must be performed as close to the real time as possible.
This entails challenges that are not present in the traditional automated classification approaches, which are usually done in some feature vector space, with an abundance of self-contained data derived from homogeneous measurements.  
In contrast, measurements generated in the synoptic sky surveys are generally sparse and heterogeneous: there are only a few initial measurements, 
their types differ from case to case, and the values have differing variances; 
the contextual information is often essential, 
and yet difficult to capture and incorporate; many sources of noise, instrumental glitches, etc., can masquerade as transient events; 
as new data arrive, the classification must be iterated dynamically.
We also require a high completeness (capture all interesting events) and a low contamination (minimize the number of false alarms). 
Since only a small fraction of the detected transient events can be followed up with the available resources, at any given stage, the current best classification should be used to make automated decisions about the follow-up priorities.
Both the classification and the availability of resources change in time, the former due to the new measurements, and the latter due to the time
allocations, weather, day/night cycle, etc.
These formidable challenges require novel approaches to a robust and flexible (near)real-time mining of massive data streams.
Reviewing the ongoing work in this domain is beyond the scope of this paper, but some examples can found in 
\cite{GSD2008,GSD2009b,GSD2010b,GSD2010c,GSD2010d,donalek2013z,richards2011a,bloom2011a,GSD2008e,GSD2011z}.

There are various free DM/KDD packages commonly used in
the academic community that would be suitable for adoption by the
astronomical community, although their uptake has also been relatively
slow. Several of them have been evaluated in this context by Donalek
et al. \cite{donalek2011}, including Orange, Rapid Miner, Weka, VoStat and DAMEWARE.

\subsection{Multidimensional Data Visualization Challenges}

Effective visualization is a key component of data exploration, analysis and understanding, and it must be an integral part of a DM process.
It is fair to say that visualization represents the bridge between the quantitative content of the data, and the intuitive understanding of it.
While astronomy can be intrinsically very visual, with images of the sky at different wavelengths and their composites, visualization of highly-
dimensional parameter spaces presents some very non-trivial challenges.
For a relevant discussion, see \cite{goodman2012}.

This is not about the images of the sky, but about a visualization of highly-dimensional parameter spaces of measurements from large sky surveys.  How do we effectively visualize phenomena that are represented in parameter spaces whose dimensionality is $D >> 3$? 
For example, a feature space of measured properties of sources in a sky survey, or a federation of several surveys, may have a dimensionality 
$D \sim 10^2 - 10^3$. 
Meaningful structures (correlations, clustering with a non-trivial topology, etc.), representing new knowledge may be present in such 
hyper-dimensional parameter spaces, and not be recognizable in any low-dimensional projection thereof.
This problem is not unique to astronomy, but it affects essentially all of "`big data"' science. 
There are fundamental limitations of the human visual perception and visual pattern recognition. 
Various tricks exist that can be used to represent up to a dozen dimensions in a pseudo-3D
graph, but going to many tens, hundreds, or thousands of dimensions
that characterize some of the modern data sets represents a fundamental
barrier to their intuitive understanding. 
This problem may be one of the key bottlenecks for data-intensive science in general.

We have experimented with a novel approach to this challenge, using an immersive virtual reality (VR) as a scientific collaboration and data visualization platform \cite{farr,donalek2014}.  
This approach offers a more intuitive perception of data and relationships present in the data (clusters, correlations, outliers, etc.), as well as a possibility of an interactive, collaborative data visualization and visual data exploration.
As the commercially-driven VR technology improves, this may become an indispensable methodology for an effective visualization of high-domensionality data spaces.

\section{DAMEWARE: A New Tool for Knowledge Discovery}
\label{sec:typestyle}

A working example of a DM/KDD platform deployed in the astronomical context is the Data Mining and Exploration Web Application REsource 
(DAMEWARE; http://dame.dsf.unina.it/) web application \cite{brescia2014b}, a joint effort between the Astroinformatics groups at University Federico II, 
the Italian National Institute of Astrophysics, and the California Institute of Technology.
DAMEWARE aims at solving in a practical way some of the above listed DM problems,
by offering a completely transparent architecture, a user-friendly interface, and the possibility to seamlessly 
access a distributed computing infrastructure. 
It adopts VO standards in order to facilitate interoperability of data; however, at the moment, it is not yet fully VO compliant.
This is partly due to the fact that new standards need to be defined for data analysis, DM methods and algorithm development. 
In practice, this implies a definition of standards in terms of an ontology and a well-defined taxonomy of functionalities to be applied to the 
astrophysical use cases.
DAMEWARE offers asynchronous access to the infrastructure tools, thus allowing the running of jobs and processes outside the scope of any 
particular web application, and independent of the user connection status.
The user, via a simple web browser, can access application resources and can keep track of his jobs by recovering related information 
(partial/complete results) without having to keep open a communication
socket. 
Furthermore, DAME has been designed to run both on a server and on a distributed computing infrastructure (e.g., Grid or Cloud).
The front end is a GUI (Graphical User Interface) that contains dynamical web pages that are used by the end users to interact 
with the applications, models, and facilities to launch scientific experiments. The
interface includes an authentication procedure that redirects the user to
a personal session environment, where he can find uploaded data, check
the experiment status and driven procedures, configure and execute new
scientific experiments. This mechanism was also required for Grid access
security reasons.
A detailed technical description of the various components can be found
in \cite{brescia2014b}. 
In the currently available DAMEWARE release, DAME offers 
Multi-Layer Perceptron ANNs, trained by three different learning rules (Back Propagation, Genetic Algorithm, Quasi-Newton), 
Random Forest and Support Vector Machines (SVM) as supervised models;  Self Organizing Feature Maps (SOFM), Principal Probabilistic Surfaces (PPS)  
and K-Means as unsupervised models.
In addition, in cases where specific problems may require applications
of DM tools or models that are not yet available in the main release,
DAME includes also a Java-based plugin wizard for custom experiment
setup (DMPlugin). This allows any user to upload into the DAME suite
their own DM code and run it on the computing infrastructure.
Even though still under development, DAME has already been tested
against several specific applications. The already mentioned globular cluster identification problem \cite{brescia2012};
the evaluation of photometric redshifts for galaxies and quasars \cite{dabrusco2007, laurino2011, brescia2013}, 
the identification of candidate quasars from multiband survey data \cite{dabrusco2009}, and
finally, the identification of candidate emission line galaxies \cite{cavuoti2014el}.
In what follows we shall summarize some of these applications.

\begin{figure}[htb]
\begin{minipage}[b]{1.0\linewidth}
\centerline{\epsfig{figure=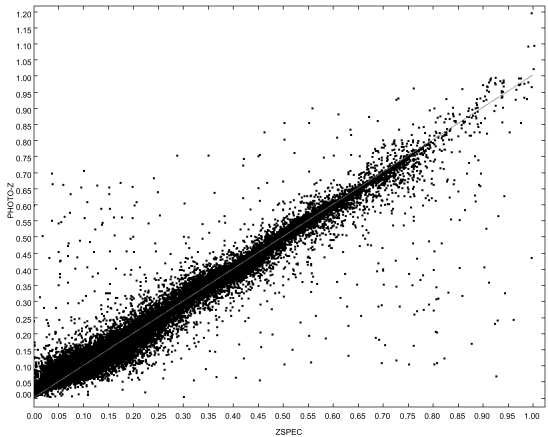,width=8.0cm}}
\caption{Distribution of spectroscopic vs photometric redhifts for the objects in the SDSS-DR9 test set.}
\end{minipage}
\end{figure}

\subsection{A use case: photometric redshifts with DAMEWARE}\label{sec:photoz}
Photo-z are essential in constraining dark matter and dark energy through weak gravitational lensing, for the identification of galaxy clusters 
and groups, for type Ia Supernovae, and to study the mass function of galaxy clusters, just to quote a few applications.

The physical mechanism responsible for the correlation between the photometric features
and the redshift of an astronomical source is the change in the contribution to the observed fluxes
caused by the prominent features of the spectrum shifting through the different filters as the spectrum of the source is redshifted. 
This mechanism implies a non-linear mapping between the photometric parameter space of
the galaxies and the redshift values. 

When accurate and multi-band photometry for a large number of objects is complemented by spectroscopic
redshifts for a statistically significant sub-sample of the same objects, the mapping function
can be inferred using supervised machine learning methods such as Neural Networks. 

As an example let us consider the recent application of the MLPQNA neural network (for details on this method see \cite{brescia2013}) 
applied to the Sloan Digital Sky Survey Data Release 9 \cite{brescia2014}. 
The SDSS-DR9 provides an ideal data set for this type of applications since it contains both very extensive multiband photometry for more than 
300 million objects as well as accurate spectroscopic redshift for a fair subsample of them. 
After an extensive series of experiments, the best results were obtained with a two hidden layer network, using a combination of the four SDSS colors 
(obtained from the SDSS $psfMag$) plus the pivot magnitude $psfMag$ in the $r$ band. 
This yields a normalized overall uncertainty of $\sigma=0.023$ with a very small average bias of $\sim 3\times 10^{-5}$, a low $NMAD$, and a low fraction of 
outliers ($\sim5\%$ at $2\sigma$ and $\sim0.1\%$ at $0.15$). 
After the rejection of catastrophic outliers, the residual uncertainty is $\sigma=0.0174$.

The trained network was then used to process the galaxies in the SDSS-DR9, matching the above outlined selection criteria, and to produce the complete photometric catalogue. 
This catalog consists of photo-z estimates for more than $143$ million SDSS-DR9 galaxies. 
The distribution of the spectroscopic versus photometric redshifts in the SDSS-DR9 test set used do derive these results is given in Fig. 2.
\section{Future Prospects}
The preceding discussion gives just a flavor of the data processing and analysis challenges in modern, data-intensive astronomy. 
We are now entering the Petascale regime in terms of data volumes, but the exponential growth continues. 
One important recent development is the advent of synoptic sky surveys, which cover large
areas of the sky repeatedly, thus escalating the challenges of data handling and analysis from massive data sets to massive data 
streams, with all of the added complexities. 
This trend is likely to continue, pushing astronomy towards the Exascale regime. 

The astronomical community has responded well and in a timely manner to the challenges of massive data handling, 
by embracing Internet-accessible archives, databases, interoperability, standard formats and
protocols, and a virtual scientific organization, Virtual Observatory, that
is now effectively a global data grid of astronomy. 
While this complex and necessary infrastructure represents a solid foundation for a big 
data science, it is just a start.
The real job of science, data analysis and knowledge discovery, starts
after all the data processing and data delivery through the archives. 
This requires some powerful new approaches to data exploration and analysis,
leading to knowledge discovery and understanding.
Many good statistical and data mining tools and methods exist, and
are gradually permeating the astronomical community, although their uptake has been slower than what may be hoped
for. 

One tangible technical problem is the scalability
of DM tools: most of the readily available ones do not scale well to the
massive data sets that are already upon us. The key problem is not so
much the data volume (expressible, e.g., as a number of feature vectors
in some data set), but their dimensionality: most algorithms may work
very well in 2 or 3 or 6 dimensions, but are simply impractical when the
intrinsic dimensionality of the data sets is measured in tens, hundred,
or thousands. 
Effective, scalable software and a methodology needed for
knowledge discovery in modern, large and complex data sets typically
do not exist yet, at least in the public domain.

\bigskip
{\bf Acknowledgments:}~~
SGD and CD acknowledge a partial support from the U.S. NSF grants AST-0834235, AST-1313422, AST-1413600, and IIS-1118041.
GL, MB and SC aknowledge a partial support from the PRIN MIUR "`The obscure universe and the evolution of baryons"' as well as
the European COST action TD-1403 "`Big data era in Sky-Earth Observations"'. 
We thank numerous colleagues for useful discussions and collaborations over the years, and in particular to 
M. Graham, A. Mahabal, A. Drake, K. Polsterer, M. Turmon, G. Riccio, D. Vinkovic, and many others.



\end{document}